\documentclass[11pt]{article} 
\usepackage{graphicx}% Include figure files 
\newcommand{\be}{\begin{equation}}
\newcommand{\ee}{\end{equation}}
\newcommand{\bea}{\begin{eqnarray}}
\newcommand{\eea}{\end{eqnarray}}
\newcommand{\sn}{{\rm sn}}

\newcommand{\dn}{{\rm dn}}
\newcommand{\cn}{{\rm cn}}
\newcommand{\sech}{{\rm sech}}

\begin{document}

\vspace{0.5in}
\begin{center}
{\LARGE{\bf Hyperbolic, Trigonometric and Periodic Solutions of 
Local and Nonlocal Fokas-Lennels Equations}}
\end{center}

\begin{center}
{\LARGE{\bf Avinash Khare}} \\
{Physics Department, Savitribai Phule Pune University \\
 Pune 411007, India}
\end{center}

\begin{center}
{\LARGE{\bf Avadh Saxena}} \\ 
{Theoretical Division and Center for Nonlinear Studies, 
Los Alamos National Laboratory, Los Alamos, New Mexico 87545, USA}
\end{center}

\vspace{0.9in}
\noindent{\bf {Abstract:}}

We obtain a large number of exact hyperbolic, trigonometric, and periodic solutions 
in terms of Jacobi elliptic functions as well as algebraic solutions with a power law 
tail of the integrable local Fokas-Lennels equation and integrable nonlocal 
Fokas-Lennels equation. Further, we consider a one-parameter family of generalized
Fokas-Lenells equations and obtain a few of their exact solutions.

\section{Introduction}

The Fokas-Lennels equation (FLE), which is an integrable equation 
\cite{fok,len1, len2} has attracted considerable attention in recent years
because of its applications in optical fibers phenomena. The most significant 
application of optical fibers is in the area of media transmission. It is worth
pointing out here that the celebrated nonlinear Schr\"odinger equation (NLSE),
which is an integrable equation, is used as a standard model for pulse 
propagation in fiber optics. However, the NLSE can only describe the 
propagation of pulses up to the first order nonlinear effect, i.e. the Kerr effect. 
In order to include certain higher order nonlinear 
effects, i.e. self-steepening, Raman scattering and third order dispersion, 
more than fifteen years back Fokas and Lennels proposed FLE, which is a general 
version of NLSE. Its physical foundation was discussed in \cite{lu}. In recent 
years FLE has been derived as a model to describe 
femtosecond pulse propagation through a single mode optical silica 
fiber. For recent progress on the study of optical solitons in fiber lasers, see
e.g. \cite{song} and references therein.

It is worth pointing out here that a long time before Fokas and Lennels, Davydova 
et al. \cite{dav1,dav2} had analyzed a FLE-like equation in some detail and 
obtained some of its solutions. Over the years, large number of solutions 
like N-soliton and breather solutions have been obtained \cite{mat1,mat2,wang,
vek,tal,dut,liu1,liu2,zha,las,zha2,zha3,ai,zhao}. It is worth pointing out that
unlike the NLS equation, FLE admits both the dark and the bright soliton solutions
without the sign change of the nonlinear term. Rogue wave solutions 
of FLE have also been constructed \cite{he,xu}. Few years back a nonlocal
FLE has been proposed and shown to be integrable \cite{zha1}. Further, in 
\cite{zha1,li,fan} the dark and bright soliton solutions, kink solutions, 
periodic solutions of the integrable nonlocal FLE have also been obtained. 

In view of the importance of FLE in the context of propagation in fiber optics,
it is important to obtain as many solutions of FLE as possible. This is the 
main motivation of the present paper. By following the ansatz proposed by 
Lennels \cite{len2} we obtain a large number ($41$ in all) of hyperbolic, 
trigonometric, and periodic solutions in terms of Jacobi elliptic functions as 
well as algebraic solutions with a power law tail of FLE. By essentially using 
a similar ansatz we show that a large number ($37$ to be precise) of exact 
solutions of the FLE are
also the solutions of the corresponding integrable nonlocal FLE as proposed 
in \cite{zha1}, most ($32$ to be precise) under the same conditions while the 
remaining five solutions under different conditions. Further, we generalize FLE
by considering a one-parameter family of FLEs. By choosing a suitable ansatz we
reduce the problem of finding the solutions of the one-parameter FLEs to that 
of finding exact solutions of the $\phi^{4n+2}$-$\phi^{2n+2}$-$\phi^2$
field theory model and obtain its four exact solutions. For the special case of
$n = 1/2$ we are able to obtain a large number of hyperbolic, trigonometric, 
and periodic solutions in terms of Jacobi elliptic functions as well as algebraic 
solutions with a power law tail.

The plan of the paper is the following. In Sec. II we set up the formalism
and following Lennels \cite{len3}, reduce the problem of obtaining the 
solutions of the integrable FLE equation 
\be\label{1}
u_{xt} = u -i\sigma |u|^2 u_{x}\,,~~\sigma = \pm 1\,,
\ee
to that of the following $\phi^6$ field equation
\be\label{2}
\phi''(y) = c_3 \phi + c_2 \phi^3 + c_1 \phi^{5} + \frac{c_4 A}{\phi^3}\,,
\ee
where $A$ is a constant of integration while $y$ to be defined below is a 
function of $x$ and $t$. We then obtain a few exact solutions of Eq. (\ref{1})
in case the constant of integration is nonzero. In Sec. III we consider 
Eq. (\ref{2}) in the special case of $A = 0$ and obtain a large number of 
its exact solutions. In particular, in Sec. IIIa we obtain the hyperbolic
solutions, in Sec. IIIb the trigonometric solutions, in Sec. IIIc the 
algebraic solutions with a power law tail and finally in Sec. IIId the periodic
solutions in terms of Jacobi elliptic functions. 
In Sec. IV we also consider an integrable nonlocal FLE \cite{zha1} 
\be\label{1a}
u_{xt} = u(x,t) -i\sigma u(-x,-t) u(x,t) u_{x}(x,t)\,,~~\sigma = \pm 1\,,
\ee
and by choosing a suitable ansatz similar to that of Eq. (\ref{2}) 
obtain its several exact hyperbolic, trigonometric, and periodic solutions in terms 
of Jacobi elliptic functions as well as algebraic solutions. In particular, we show 
that while many of the  solutions of the 
local FLE obtained in Sec. II and III are also the solutions of the nonlocal 
FLE under the same conditions on the parameters, there are a few solutions of the
local FLE obtained in Sec. III that are also the
solutions of the nonlocal FLE but under different conditions on the parameters.
Besides, there are four solutions of the local FLE obtained in Sec. III which 
are not the solutions of the nonlocal FLE (\ref{1a}). In Sec. V we consider a 
one-parameter family of FLEs given by 
\be\label{1b}
u_{xt} = u -i\sigma |u|^{2n} u_{x}\,,~~\sigma = \pm 1\,,~~n = 1/2, 1, 3/2, 
2,... \,,
\ee
and by choosing a suitable ansatz reduce the solutions to that of the following
field theory equation
\be\label{1c}
\phi''(y) = b_3 \phi + b_2 \phi^{2n+1} +b_1 \phi^{4n+1} +\frac{b_4 A}{\phi}
+\frac{b_5}{\phi^3}\,,
\ee
where $A$ is a constant of integration. We then obtain four exact solutions of this
equation. In the Appendix we consider the special case of $n = 1/2$ which 
corresponds to a FLE related to the quadratic NLSE. Finally, in Sec. VI we 
summarize the main results obtained in the paper and point out some of the 
possible open problems. In the Appendix we obtain a large number of exact 
solutions of Eq. (\ref{1c}) in the special case of $n = 1/2$ which is related 
to the quadratic NLSE.

\section{Suitable Ansatz For Finding Solutions of FLE}

Following Lennels \cite{len3}, we start with the ansatz
\be\label{3}
u(x,t) = \phi(y) e^{i[k x-\omega t +\theta(y)]}\,,~~y = x -vt\,.
\ee
On substituting the ansatz (\ref{3}) in FLE (\ref{1}) and separating real and 
imaginary parts we get two equations. The real part gives the equation
\be\label{4}
v\phi''(y) = - \sigma [k+\theta'(y)] \phi^3(y) +(k\Omega-1)\phi(y) 
+v [\theta'(y)]^2 \phi(y) +(\Omega + vk) \theta'(y) \phi(y)\,.
\ee
On the other hand, the imaginary part gives the equation
\be\label{5}
\sigma \phi^2(y)\phi'(y) = v\phi(y)\theta''(y) +[\Omega +kv +2v\theta'(y)]
\phi(y)\,.
\ee
On multiplying Eq. (\ref{5}) by $\phi(y)$ and integrating we obtain
\be\label{6}
\theta'(y) = \frac{\sigma \phi^2}{4v} -\frac{(\Omega+kv)}{2v} 
+ \frac{A}{v \phi^2}\,,
\ee
where $A$ is constant of integration. On substituting this expression for 
$\theta'(y)$ in Eq. (\ref{4}) for the real part, we find that $\phi(y)$-like
satisfies the $\phi^6$ equation
\be\label{7}
\phi''(y) = c_1 \phi^5 + c_2 \phi^3 + c_3 \phi + c_4 \phi^{-3}\,,
\ee
where
\bea\label{8}
&&c_1 = -\frac{3}{16v^2} < 0\,,~~ c_2 = \sigma \frac{(\omega -vk)}{2v^2}\,,
\nonumber \\
&&c_3 = -\frac{(vk-\Omega)^2 +2A\sigma+4v}{4v^2}\,,~~
c_4 = \frac{A^2}{v^2} > 0\,.
\eea
Note that the only $\sigma$ dependent terms are there in case $A \ne 0$ since
$\sigma^2 = 1$. Further, while $c_1 < 0$, $c_4 > 0$, but $c_2$, $c_3$ could be
zero, negative or positive depending on the values of the parameters $v,k,
\Omega, A$. In this section we present the solution of the $\phi^6$-like 
Eq. (\ref{7}) in case the constant of integration $A \ne 0$ while in the next
section we present solutions in case the constant of integration $A = 0$.

\subsection{Solutions when Constant of Integration $A \ne 0$}

We now consider the general case when the constant of integration $A$ is 
nonzero and present seven solutions of Eq. (\ref{7}) with $c_1$, $c_2$, $c_3$, $c_4$ 
being given by Eq. (\ref{8}) \cite{kody}. 

{\bf Solution I}

It is easy to check that one of the periodic pulse 
solutions to Eqs. (\ref{7}) and (\ref{8}) is 
\be\label{1.3}
\phi = \sqrt{D\cn(\beta y,m)+B}\,, ~~B > D > 0\,,
\ee
provided
\bea\label{1.4}
&&c_1 D^2 = -\frac{3m\beta^2}{4}\,,~~c_2 D^2 = 2m B \beta^2\,,~~c_3 = 
\frac{(2m-1)\beta^2}{4} - \frac{9 c_{2}^{2}}{32|c_1|}\,, \nonumber \\
&&c_4 = \frac{27 c_{2}^{4}}{4096|c_1|^3} -\frac{9(2m-1)c_{2}^{2} \beta^2}
{256 |c_1|^2} -\frac{3m(1-m)\beta^4}{16|c_1|}\,.
\eea
Thus for this solution $c_1 < 0$, $c_2 > 0$. On demanding that $c_4 > 0$ gives a 
bound on $\beta^2$ in terms of $c_2$, $c_1, m$.

{\bf Solution II}

Another periodic pulse solution to Eqs. (\ref{7}) and (\ref{8}) is
\be\label{1.5}
\phi = \sqrt{D\dn(\beta y,m)+B}\,, ~~B, D > 0\,,
\ee
provided
\bea\label{1.6}
&&c_1 D^2 = -\frac{3\beta^2}{4}\,,~~c_2 D^2 = 2 B \beta^2\,,~~c_3 = 
\frac{(2-m)\beta^2}{4} - \frac{9 c_{2}^{2}}{32|c_1|}\,, \nonumber \\
&&c_4 = \frac{27 c_{2}^{4}}{4096|c_1|^3}-\frac{9(2-m) c_{2}^{2} \beta^2}
{256 |c_1|^2} +\frac{3(1-m)\beta^4}{16|c_1|} > 0\,.
\eea
Thus for this solution also $c_1 < 0, c_2 > 0$. On demanding that $c_4 > 0$ 
gives a bound on $\beta^2$ in terms of $c_2$, $c_1, m$.

{\bf Solution III}

In the limit $m = 1$, both solutions I and II go over to the hyperbolic pulse
solution
\be\label{1.7}
\phi = \sqrt{D\sech(\beta y)+B}\,, ~~B, D > 0\,,
\ee
provided
\bea\label{1.8}
&&c_1 D^2 = -\frac{3\beta^2}{4}\,,~~c_2 D^2 = 2 B \beta^2\,,~~c_3 = 
\frac{\beta^2}{4} - \frac{9 c_{2}^{2}}{32|c_1|}\,, \nonumber \\
&&c_4 = \frac{27 c_{2}^{4}}{4096|c_1|^3} -\frac{9 b^2 \beta^2}{256 |c_1|^2}\,. 
\eea
Thus for this solution also $c_1 < 0$, $c_2 > 0$. On demanding that $c_4 > 0$ 
gives a bound 
\be\label{1.9}
\beta^2 < \frac{3c_{2}^{2}}{16 |c_1|}\,.
\ee

{\bf Solution IV}

It is easy to check that another periodic pulse solution to 
Eqs. (\ref{7}) and (\ref{8}) is
\be\label{1.10}
\phi = \frac{1}{\sqrt{D\cn(\beta y,m)+B}}\,, ~~B > D > 0\,,
\ee
provided
\bea\label{1.11}
&&D = 1\,,~~c_4 = \frac{m\beta^2}{4}\,,~~c_2 = [2m B^2-(2m-1)] B\beta^2\,,
\nonumber \\
&&c_3 = [(2m-1)-6 B^2]\frac{\beta^2}{4}\,,~~c_1 = -[m B^4 -(2m-1) B^2 
-(1-m)]\frac{3\beta^2}{4}\,.  \nonumber \\ 
\eea
Thus for this solution $c_4$, $c_2 > 0$ while $c_1$, $c_3 < 0$ since $B > 1$.

{\bf Solution V}

Another periodic pulse solution to Eqs. (\ref{7}) and (\ref{8}) is
\be\label{1.12}
\phi = \frac{1}{\sqrt{D\dn(\beta y,m)+B}}\,, ~~D, B > 0\,,
\ee
provided
\bea\label{1.13}
&&D = 1\,,~~c_4 = \frac{\beta^2}{4}\,,~~c_2 = [2 B^2-(2-m)] B\beta^2\,,
\nonumber \\
&&c_3 = [(2-m)-6 B^2]\frac{\beta^2}{4}\,,~~c_1 = -[B^4 -(2-m) B^2 
+(1-m)]\frac{3\beta^2}{4}\,. \nonumber \\ 
\eea
Thus for this solution too $c_4$, $c_2 > 0$ while $c_1$, $c_3 < 0$ since $B > 1$.

{\bf Solution VI}

In the limit $m = 1$ both the solutions IV and V go over to the hyperbolic 
pulse solution
\be\label{1.14}
\phi = \frac{1}{\sqrt{D\sech(\beta y)+B}}\,, ~~D, B > 0\,,
\ee
provided
\bea\label{1.15}
&&D = 1\,,~~c_4 = \frac{\beta^2}{4}\,,~~c_2 = [2 B^2- 1] B\beta^2\,,
\nonumber \\
&&c_3 = [1-6 B^2]\frac{\beta^2}{4}\,,~~c_1 = -B^2 (B^2 -1) \frac{3\beta^2}{4}\,.
\eea
Thus for this solution too $c_4$, $c_2 > 0$ while $c_1$, $c_3 < 0$ since $B > 1$.

{\bf Solution VII}

Remarkably, Eqs. (\ref{7}) and (\ref{8}) also admit an algebraic solution 
with a power law tail
\be\label{1.16}
\phi = \sqrt{\frac{E^2 y^2 + B^2}{D^2+ y^2}}\,,~~B^2 \ne D^2 E^2\,,
\ee
provided
\bea\label{1.17}
&&c_4 = \frac{E^6 B^2}{(B^2-D^2 E^2)^2}\,,~~ c_2 = 2\frac{(B^2+3E^2 D^2)}
{(B^2-E^2 D^2)^2}\,, \nonumber \\
&&c_1 = -\frac{3D^2}{(B^2 - E^2 D^2)^2}\,,~~ c_3 = -\frac{3E^2(B^2+E^2 D^2)}
{(B^2-E^2 D^2)^2}\,. 
\eea

\section{Exact Solutions of Eqs. (\ref{7}) and (\ref{8}) in Case $A = 0$}

We now consider exact solutions of Eqs. (\ref{7}) and (\ref{8}) in case the
constant of integration $A = 0$. In case $A = 0$, from Eqs. (\ref{7})
and (\ref{8}) we find that we need to solve a simpler equation
\be\label{2.1}
\phi''(y) = -\frac{3}{16v^2} \phi^5 +\frac{\sigma(\Omega -vk)}{2v^2} \phi^3
-\frac{(vk-\Omega)^2+ 4v}{4v^2} \phi\,.
\ee
Notice that while $\phi^5$ coefficient $c_1$ is always negative, the 
coefficient of $\phi^3$, i.e. $c_2$ is negative (positive) depending on if 
$\sigma(\omega-vk) < (>) $ 0. Note $\sigma = \pm 1$. On the other hand the
coefficient of $\phi$, i.e. $c_3$ is negative (positive) depending on 
whether $(vk-\Omega)^2 +4v > (<)$ 0. 

Remarkably, we find that Eq. (\ref{2.1}) admits a large number of solutions, 
hyperbolic, trigonometric, Jacobi elliptic periodic pulse and kink 
solutions as well as kink and pulse solutions with a power law tail \cite{snb,
falk,avadh,ks1}. We 
mention these solutions one by one. In each case we also discuss the 
possibility of having solutions with $c_3 = 0$, i.e. when $\Omega = vk$, as 
well as the possibility of solutions with $c_2 = 0$, i.e. when $(vk-\Omega)^2 = 
-4v$ which is only possible if $v < 0$.

\subsection{Hyperbolic Solutions}

We now present seven hyperbolic solutions to Eq. (\ref{2.1}).

{\bf Solution I}

It is easy to check that
\be\label{2.2}
\phi(y) = \frac{A}{\sqrt{B+\cosh(\beta y)}}\,,~~ -1 < B < 1\,,
\ee
is an exact solution of Eq. (\ref{2.1}) provided
\be\label{2.3}
c_1 A^4 = -\frac{3(1-B^2)\beta^2}{4} < 0\,,~~c_2 A^2 = -B \beta^2\,,~~
c_3 = \frac{\beta^2}{4} > 0\,.
\ee
Thus this is an acceptable solution if $-1 < B < 1$ and in that case 
$c_1 < 0$, $c_3 > 0$ while $c_2 > (<)$ 0 depending on if $B < (>)$ 0. 

{\bf Solution II}

In the special case of $B = 0$ we then have the solution
\be\label{2.4}
\phi(y)  = A \sech^{1/2}(\beta y)\,,
\ee
provided
\be\label{2.5}
c_1 A^4 = -\frac{3\beta^2}{4} < 0\,,~~c_2 = 0\,,~~ 
c_3 = \frac{\beta^2}{4} > 0\,.
\ee

{\bf Solution III}

Another solution to Eq. (\ref{2.1}) is
\be\label{2.6}
\phi = A \sqrt{1+\sech(\beta y)}\,,
\ee
provided
\be\label{2.7}
c_3 = -\frac{5\beta^2}{4}\,,~~c_2 A^2 = 2\beta^2\,,~~c_3 A^4 
= -\frac{3\beta^2}{4}\,.
\ee
Thus for this solution $c_1$, $c_3 < 0$, $c_2 > 0$. 

{\bf Solution IV}

Another solution to Eq. (\ref{2.1}) is
\be\label{2.8}
\phi(x) = A \sqrt{\frac{1-\sech(\beta y)}{B+\sech(\beta y)}}\,,~~B > 1\,,
\ee
provided
\bea\label{2.9}
&&(1+B)c_3 = -\frac{(5B-1)\beta^2}{4}\,,~~(1+B) c_2 A^2 = B(2B-1) \beta^2\,,
\nonumber \\
&&(1+B) c_1 A^4 = -\frac{3(B-1)B^2}{4} \beta^2\,.
\eea
Thus for this solution $c_1$, $c_2$, $c_3 < 0$ in case $B > 1$.

{\bf Solution V}

Another solution to Eq. (\ref{2.1}) is
\be\label{2.10}
\phi(x) = A \sqrt{\frac{1+\sech(\beta y)}{B+\sech(\beta y)}}\,,~~ B > 1\,,
\ee
provided
\bea\label{2.11}
&&(B-1)c_3 = -\frac{(5B+1)\beta^2}{4}\,,~~(B-1) c_2 A^2 = B(2B+1) \beta^2\,,
\nonumber \\
&&(B-1) c_1 A^4 = -\frac{3(B+1)B^2}{4} \beta^2\,.
\eea
Thus for this solution $c_1$, $c_3 < 0$, $c_2 > 0$ in case $B > 1$.

{\bf Solution VI}

Eq. (\ref{2.1}) admits a (hyperbolic) kink solution
\be\label{2.12}
\phi = \frac{A\sinh(\beta y)}{\sqrt{B+\cosh^2(\beta y)}}\,,~~ -1 < B < 0 \,,
\ee
provided
\be\label{2.13}
(1-|B|) c_3 = -(|B|+2)\beta^2\,,~~(1-|B|)c_2 A^2 = (2|B|+1)\beta^2\,,~~
c_1 A^4 = -3|B| \beta^2\,.
\ee
Thus for this solution $c_1$, $c_3 < 0$, $c_2 > 0$ in case $-1 < B < 0$.

{\bf Solution VII}

Yet another hyperbolic solution to Eq. (\ref{2.1}) is 
\be\label{2.14}
\phi = \frac{A\cosh(\beta y)}{\sqrt{B+\cosh^2(\beta y)}}\,,~~ -1 < B < 0\,,
\ee
provided
\be\label{2.15}
|B|c_3 = -(3-|B|)\beta^2\,,~~|B| c_2 A^2 = 2(3-2|B|)\beta^2\,,~~
|B|cA^4 = -3(1-|B|)\beta^2\,.
\ee
Thus for this solution  too $c_1$, $c_3 < 0$, $c_2 > 0$ in case $-1 < B < 0$.

{\bf Trigonometric Solutions}

We now present five trigonometric solutions, four periodic pulse and one 
periodic kink solution.

{\bf Solution VIII}

It is easily checked that
\be\label{3.1}
\phi = \frac{A}{1+B\cos^2(\beta y)}\,,~~ B > 0\,,
\ee
is a periodic pulse solution of Eq. (\ref{2.1}) provided
\be\label{3.2}
B c_3 = - \beta^2\,,~~B c_2 A^2 = 2(B+2)\beta^2\,,~~B c_1 A^4 
= -3(B+1)\beta^2\,.
\ee
Thus for this solution while $c_1$, $c_3 < 0$, $c_2 > 0$ when $B > 0$.

{\bf Solution IX}

Another trigonometric pulse solution to Eq. (\ref{2.1}) is
\be\label{3.3}
\phi = A \sqrt{\frac{1+\cos(\beta y)}{B+\cos(\beta y)}}\,,~~B > 1\,,
\ee
provided
\be\label{3.4}
(B-1)c_3 = -\frac{(B+5)\beta^2}{4}\,,~~(B-1)c_2 A^2 = (B+2)\beta^2\,,
~~(B-1) c A^4 = -\frac{3}{4} (1+B)\beta^2\,.
\ee
Thus for this solution $c_1$, $c_3 < 0$, $c_2 > 0$ in case $B > 1$.

{\bf Solution X}

Another trigonometric pulse solution to Eq. (\ref{2.1}) is
\be\label{3.5}
\phi = A \sqrt{\frac{1-\cos(\beta y)}{B+\cos(\beta y)}}\,,~~B > 1\,,
\ee
provided
\be\label{3.6}
(B+1)c_3 = -\frac{(5-B)\beta^2}{4}\,,~~(B+1)c_2 A^2 = -(B-2)\beta^2\,,
~~(B+1) c A^4 = -\frac{3}{4} (B-1)\beta^2\,.
\ee
Thus for this solution $c_1 < 0$ while $c_3 = 0$ if $B = 5$ while 
$c_3 > (<)$ 0 depending on if $B > (<) $ 5. On the other hand, $c_2 = 0$
if $B = 2$ while $c_2 > (<)$ 0 depending on if $B < (>)$ 2. Note that 
one always has $B > 1$ so that $c_1 < 0$.

{\bf Solution XI}

Another trigonometric pulse solution to Eq. (\ref{2.1}) is
\be\label{3.7}
\phi = \frac{A\cos(\beta y)}{\sqrt{1+B\cos^2(\beta y)}}\,,~~B > 0\,,
\ee
provided
\be\label{3.8}
c_3 = -(3B+1)\beta^2\,,~~c_2 A^2 = 2B(2+3B)\beta^2\,,
~~c_1 A^4 = -3B^2(B+1)\beta^2\,.
\ee
Notice that for this solution $c_1$, $c_3 < 0$, $c_2 > 0$ in case $B > 0$.

{\bf Solution XII}

Eq. (\ref{2.1}) also admits the trigonometric kink solution
\be\label{3.9}
\phi = \frac{A\sin(\beta y)}{\sqrt{1+B\cos^2(\beta y)}}\,,
~~B > 1\,,
\ee
provided
\bea\label{3.10}
&&(B+1)c_3 = (2B-1)\beta^2\,,~~(B+1)c_2 A^2 = -2B(2-B)\beta^2\,, \nonumber \\
&&(B+1) c_1 A^4 = -3 B (B-1) \beta^2\,.
\eea
Thus for this solution while $c_1 < 0$, $c_3 > 0$, $c_2 = 0$ in case $B = 2$ 
while $c_2 > 0$ if $B > 2$. On the other hand, if $1 < B < 2$ then $c_2 < 0$.

\subsection{Solutions With Power Law Tail}

We now present four solutions with a power law tail, three pulse solutions and 
one kink solution. 

{\bf Solution XIII}

It is easy to check that Eq. (\ref{2.1}) admits the pulse solution
\be\label{3.11}
\phi = \frac{A}{\sqrt{B+y^2}}\,,~~ B > 0\,,
\ee
with a power law tail provided
\be\label{3.12}
c_3 = 0\,,~~c_2 A^2 = 2\,,~~ c_1 A^4 = -3B\,.
\ee
Thus for this solution while $c_1 < 0$, $c_2 > 0$ but $c_3 = 0$.

{\bf Solution XIV}

Eq. (\ref{2.1}) also admits the kink solution 
\be\label{3.13}
\phi = \frac{Ay}{\sqrt{B+y^2}}\,,~~ B > 0\,,
\ee
with a power law tail provided
\be\label{3.14}
B c_3 = -3\,,~~B c_2 A^2 = 6\,,~~ B c_1 A^4 = -3\,.
\ee
Thus for this solution while $c_1$, $c_3 < 0$, $c_2 > 0$.

{\bf Solution XV}

Another pulse solution with a power law tail to Eq. (\ref{2.1}) is
\be\label{3.15}
\phi = \frac{A}{B+y^2}\,,~~ B > 0\,,
\ee
with a power law tail provided
\be\label{3.16}
c_3 = 0\,,~~c_2 A^2 = 6\,,~~ c_1 A^4 = -8B\,.
\ee
Thus for this solution while $c_1 < 0$, $c_2 > 0$ but $c_3 = 0$.

{\bf Solution XVI}

Eq. (\ref{2.1}) also admits the pulse solution with a power law tail
\be\label{3.17}
\phi = \frac{Ay}{B+y^2}\,,~~ B > 0\,,
\ee
provided
\be\label{3.18}
B c_3 = -6\,,~~B c_2 A^2 = 14\,,~~ B c_1 A^4 = -8\,.
\ee
Thus for this solution while $c_1$, $c_3 < 0$, $c_2 > 0$.

\subsection{Periodic Solutions in Terms of Jacobi Elliptic Functions}

We now present a large number of periodic solutions of Eq. (\ref{2.1}) in terms
of Jacobi elliptic functions.

{\bf Solution XVII}

It is easily shown that 
\be\label{4.1}
\phi = A \sqrt{1+\cn(\beta y,m)}\,,
\ee
is an exact periodic pulse solution of the field Eq. (\ref{2.1}) provided
\be\label{4.2}
c_3 = -\frac{(4m+1)\beta^2}{4}\,,~~ c_2 A^2 = -2 m \beta^2\,,~~ c_1 A^4 =
= -\frac{3m\beta^2}{4}\,.
\ee
Thus for this solution $c_1$, $c_2$, $c_3 < 0$. 

{\bf Solution XVIII}

It is easily shown that 
\be\label{4.3}
\phi = A \sqrt{1+\dn(\beta y,m)}\,,
\ee
is an exact periodic pulse solution of the field Eq. (\ref{2.1}) provided
\be\label{4.4}
c_3 = -\frac{(4+m)\beta^2}{4}\,,~~c_2 A^2 = -2 \beta^2\,,~~ c_1 A^4 = 
-\frac{3\beta^2}{4}\,.
\ee
Thus for this solution $c_1$, $c_2$, $c_3 < 0$. 

{\bf Solution XIX}

It is easily shown that 
\be\label{4.5}
\phi = A \sqrt{\dn(\beta y,m)+ \sqrt{1-m}}\,,
\ee
is an exact periodic pulse solution of the field Eq. (\ref{2.1}) provided
\be\label{4.6}
c_3 = \frac{(5m-4)\beta^2}{4}\,,~~ c_2 A^2 = -2\sqrt{1-m}\beta^2\,,~~c_1 A^4
= -(3/4)\beta^2\,.
\ee
Thus for this solution $c_1$, $c_2 < 0$ while $c_3 = 0$ in case $m = 4/5$. On the
other hand, $c_3 > 0$ if $4/5 < m < 1$ while $c_3 < 0$ if $0 < m < 4/5$.

{\bf Solution XX}

It is easily shown that 
\be\label{4.7}
\phi = A \sqrt{\dn(\beta y,m)- \sqrt{1-m}}\,,
\ee
is an exact periodic pulse solution of the field Eq. (\ref{2.1}) provided
\be\label{4.8}
c_3 = \frac{(5m-4)\beta^2}{4}\,,~~c_2 A^2 = 2\sqrt{1-m}\beta^2\,,~~c_1 A^4
= -(3/4)\beta^2\,.
\ee
Thus for this solution $c_1$, $c_2 < 0$ while $c_3 = 0$ in case $m = 4/5$. On the
other hand, $c_3 > 0$ if $4/5 < m < 1$ while $c_3 < 0$ if $0 < m < 4/5$.

{\bf Solution XXI}

It is easy to check that 
\be\label{4.9}
\phi = A \sqrt{\frac{1+\dn(\beta y,m)}{B+\dn(\beta y,m)}}\,,
\ee
is an exact periodic pulse solution of Eq. (\ref{2.1}) provided
\bea\label{4.10}
&&(B-1)c_3 = -[(4+m)B+(5m-4)]\frac{\beta^2}{4}\,,~~(B-1) c_2 A^2 
= [B(2B+m)-2(1-m)]\beta^2\,, \nonumber \\
&&(B-1) c_1 A^4 = -\frac{3}{4} (1+B)[B^2-(1-m)]\beta^2\,.
\eea
Thus for this solution $c_1 < 0$ in case either $B > 1$ or $ B < \sqrt{1-m}$.
Note that for this solution $c_3 = 0$ in case $B = \frac{(4-5m)}{(4+m)}$ 
while $c_2 = 0$ in case $B = \frac{\sqrt{m^2+16(1-m)}-m}{4}$.

{\bf Solution XXII}

It is easy to check that 
\be\label{4.11}
\phi = A \sqrt{\frac{1+\cn(\beta y,m)}{B+\cn(\beta y,m)}}\,,~~B > 1\,,
\ee
is an exact periodic pulse solution of Eq. (\ref{2.1}) provided
\bea\label{4.12}
&&(B-1)c_3 = -[(4m+1)B+(5-4m)]\frac{\beta^2}{4}\,,~~(B-1)c_2 A^2 
= [B(2m B+1)+2(1-m)]\beta^2\,, \nonumber \\
&&(B-1) c_1 A^4 = -\frac{3}{4} (1+B)[m B^2+(1-m)]\beta^2\,.
\eea
Thus for this solution $c_1$, $c_3 < 0$, $c_2 > 0$ since $B > 1$.

{\bf Solution XXIII}

It is easy to check that 
\be\label{4.13}
\phi = A \sqrt{\frac{1-\dn(\beta y,m)}{B+\dn(\beta y,m)}}\,,~~B > 1\,,
\ee
is an exact solution of Eq. (\ref{2.1}) provided
\bea\label{4.14}
&&(1+B)c_3 = -[(4+m)B+(4-5m)]\frac{\beta^2}{4}\,,~~(1+B) c_2 A^2 
= [B(2B-m)-2(1-m)]\beta^2\,, \nonumber \\
&&(1+B) c_1 A^4 = -\frac{3}{4} (B-1)[B^2+(1-m)]\beta^2\,.
\eea
Thus for this solution $c_1$, $c_3 < 0$, $c_2 > 0$ since $B > 1$.

{\bf Solution XXIV}

It is easy to check that 
\be\label{4.15}
\phi = A \sqrt{\frac{1-\cn(\beta y,m)}{B+\cn(\beta y,m)}}\,,~~ B > 1\,,
\ee
is an exact solution of Eq. (\ref{2.1}) provided
\bea\label{4.16}
&&(1+B)c_3 = -[(4m+1)B-(5-4m)]\frac{\beta^2}{4}\,,~~(1+B)c_2 A^2 = [B(2m B-1)
+2(1-m)]\beta^2\,, \nonumber \\
&&(1+B) c_1 A^4 = -\frac{3}{4} (B-1)[m B^2+(1-m)]\beta^2\,.
\eea
Thus for this solution $c_1 < 0$, $c_2 > 0$ while $c_3 = 0$ in case 
$B = \frac{(5-4m)}{1+4m}$. Note that $B > 1$ is ensured if $0 < m < 1/2$.

{\bf Solution XXV}

It is easy to check that 
\be\label{4.17}
\phi = A \sqrt{\frac{1+\sn(\beta y,m)}{B+\sn(\beta y,m)}}\,,~~ 
1 < B < 1/\sqrt{m}\,,
\ee
is an exact solution of Eq. (\ref{2.1}) provided
\bea\label{4.18}
&&(B-1)c_3 = [(5m-1)B-(5-m)]\frac{\beta^2}{4}\,,~~(B-1)c_2 A^2 
= -[m B(2B+1)-(B+2)] \beta^2\,, \nonumber \\
&&(B-1) c_1 A^4 = -\frac{3}{4} (1+B)(1-m B^2)\beta^2\,.
\eea
For this solution while $c_1 < 0$, the values of $c_2$, $c_3$ depend on
the values of $m$ and $B$.

{\bf Solution XXVI}

It is easy to check that 
\be\label{4.19}
\phi = A \sqrt{\frac{1-\sn(\beta y,m)}{B+\sn(\beta y,m)}}\,,~~ 
1 < B < 1/\sqrt{m}\,,
\ee
is an exact solution of Eq. (\ref{2.1}) provided
\bea\label{4.20}
&&(1+B)c_3 = [(5m-1)B+(5-m)]\frac{\beta^2}{4}\,,~~(1+B)c_2 A^2 = 
-[2m B^2 +B(1-m)-2]\beta^2\,, \nonumber \\
&&(1+B) c_1 A^4 = -\frac{3}{4} (B-1)(1-m B^2)\beta^2\,.
\eea 
For this solution while $c_1 < 0$, the values of $c_2$, $c_3$ depend on
the values of $m$ and $B$.

{\bf Solution XXVII}

It is easy to check that 
\be\label{4.21}
\phi = A \sqrt{\frac{\dn(\beta y,m)+\sqrt{1-m}}{B+\dn(\beta y,m)}}\,,
\ee
is an exact solution of Eq. (\ref{2.1}) provided
\bea\label{4.22}
&&(B-\sqrt{1-m})c_3 = [(5m-4)B +(4+m)\sqrt{1-m}]\frac{\beta^2}{4}\,, 
\nonumber \\
&&(B-\sqrt{1-m}) c_2 A^2 = -[m B - 2(B^2-1)\sqrt{1-m}] \beta^2\,, \nonumber \\
&&(B-\sqrt{1-m}) c_1 A^4 = -\frac{3}{4} (B+\sqrt{1-m})(B^2 -1)\beta^2\,.
\eea
For this solution $c_1 < 0$ in case either $B < \sqrt{1-m}$ or $B > 1$. In case
$B > 1$, $c_3 = 0$ in case $B = \frac{(4+m)\sqrt{1-m}}{4-5m}$ for a suitable 
range of values of $m$. Similarly, in case $B > 1, c_2 = 0$ in case
$B = \frac{m+\sqrt{m^2+16(1-m)}}{4\sqrt{1-m}}$ for a suitable range of values of 
$m$.

{\bf Solution XXVIII}

It is easy to check that 
\be\label{4.23}
\phi = A \sqrt{\frac{\dn(\beta y,m)-\sqrt{1-m}}{B+\dn(\beta y,m)}}\,,
~~B > 1\,,
\ee
is an exact solution of Eq. (\ref{2.1}) provided
\bea\label{4.24}
&&(B+\sqrt{1-m})c_3 = [(5m-4)B -(4+m)\sqrt{1-m}]\frac{\beta^2}{4}\,,
\nonumber \\
&&(B+\sqrt{1-m}) c_2 A^2 = -[2(B^2-1)\sqrt{1-m}+mB] \beta^2\,, \nonumber \\
&&(B+\sqrt{1-m}) c_1 A^4 = -\frac{3}{4} (B-\sqrt{1-m})(B^2-1)\beta^2\,.
\eea
For this solution $c_1 < 0$ in case either $B < \sqrt{1-m}$ or $B > 1$. 
In case $B > 1$, $c_3 = 0$ provided $B = \frac{(4+m)\sqrt{1-m}}{5m-4}$ and
$m > 4/5$.

{\bf Solution XXIX}

It is easy to check that 
\be\label{4.25}
\phi = A \sqrt{\frac{1+\sqrt{m}\sn(\beta y,m)}{B+\sn(\beta y,m)}}\,,
~~ 1 <B < 1/\sqrt{m}\,,
\ee
is an exact solution of Eq. (\ref{2.1}) provided
\bea\label{4.26}
&&(1-\sqrt{m}B)c_3 = -[(5-m)\sqrt{m}B -(5m-1)]\frac{\beta^2}{4}\,, 
\nonumber \\
&&(1-\sqrt{m} B)c_2 A^2 = [2\sqrt{m} (B^2-1) +(1-m)B]\beta^2\,, \nonumber \\
&&(1-\sqrt{m}B) c_3 A^4 = -\frac{3}{4} (\sqrt{m}B+1)(B^2-1)\beta^2\,.
\eea
For this solution while $c_1 < 0$, the values of $c_2$, $c_3$ depend on
the values of $m$ and $B$.

{\bf Solution XXX}

It is easy to check that 
\be\label{4.27}
\phi = A \sqrt{\frac{1-\sqrt{m}\sn(\beta y,m)}{B+\sn(\beta y,m)}}\,,
~~ 1 < B < 1/\sqrt{m}\,,
\ee
is an exact solution of Eq. (\ref{2.1}) provided
\bea\label{4.28}
&&(\sqrt{m}B+1)c_3 = [(5-m)\sqrt{m}B +(5m-1)]\frac{\beta^2}{4}\,,
\nonumber \\
&&(\sqrt{m} B+1)c_2 A^2 = -[2\sqrt{m} (B^2-1) -(1-m)B]\beta^2\,, \nonumber \\
&&(\sqrt{m}B+1) c_1 A^4 = -\frac{3}{4} (1-\sqrt{m}B)(B^2-1)\beta^2\,.
\eea
For this solution while $c_1 < 0$, the values of $c_2$, $c_3$ depend on
the values of $m$ and $B$.

{\bf Solution XXXI}

It is easy to check that
\be\label{4.29}
\phi = \frac{A\sn(\beta y,m)}{\sqrt{1+B\cn^2(\beta y,m)}}\,,
~~B > \frac{1}{1-m}\,,
\ee
is an exact periodic kink solution of Eq. (\ref{2.1}) provided
\bea\label{4.30}
&&(B+1) c_3 = [(2-m)B-(1+m)]\beta^2\,,~~(B+1)c_2 A^2 
= -2[(2B-m) -(1-m)B^2]\beta^2\,, \nonumber \\
&&c_1 A^4 = -3B[(1-m)B-1] \beta^2\,.
\eea
For this solution while $c_1 < 0$, the values of $c_2$, $c_3$ depend on
the values of $m$ and $B$.

{\bf Solution XXXII}

It is easy to check that
\be\label{4.31}
\phi = \frac{A\cn(\beta y,m)}{\sqrt{1+B\cn^2(\beta y,m)}}\,,
~~B > \frac{m}{1-m}\,,
\ee
is an exact periodic kink solution of Eq. (\ref{2.1}) provided
\bea\label{4.32}
&&c_3 = [2m-1- 3B(1-m)]\beta^2\,,~~c_2 A^2 
= -2[m +2B(2m-1) -3(1-m)B^2]\beta^2\,, \nonumber \\
&&c_1 A^4 = -3B(B+1)[(1-m)B-m]\beta^2\,.
\eea
For this solution while $c_1 < 0$, the values of $c_2$, $c_3$ depend on
the values of $m$ and $B$.

{\bf Solution XXXIII}

It is easy to check that
\be\label{4.33}
\phi = \frac{A\dn(\beta y,m)}{\sqrt{1+B\cn^2(\beta y,m)}}\,,
~~B > \frac{m}{1-m}\,,
\ee
is an exact periodic kink solution of Eq. (\ref{2.1}) provided
\bea\label{4.34}
&&[(1-m)B-m]c_3 = -[m(2-m)+(1-m^2)B]\beta^2\,, \nonumber \\
&&[(1-m)B-m] c_2 A^2 = 2[m +2B +(1-m)B^2]\beta^2\,,
\nonumber \\
&&[(1-m)B-m] m^2 c_1 A^4 = -B[3m^2+(2+4m-3m^2)B]\beta^2\,.
\eea
Notice that for this solution $c_1$, $c_3 < 0$, $c_2 > 0$  since 
$B > \frac{m}{1-m}$.

{\bf Solution XXXIV}

It is easy to check that
\be\label{4.35}
\phi = \frac{A}{\sqrt{1+B\cn^2(\beta y,m)}}\,,~~ -1 < B < 0\,,
\ee
is an exact periodic kink solution of Eq. (\ref{2.1}) provided
\bea\label{4.36}
&&|B| c_3 = -[3m-(2m-1)|B|]\beta^2\,,~~|B| c_2 A^2 = 
2[3m -2(2m-1)|B|]\beta^2\,, \nonumber \\
&&|B| c_1 A^4 = -[3m+3(2m-1)|B|+2(1-m)|B|^3]\beta^2\,.
\eea
Notice that for this solution $c_1$, $c_2$, $_3 < 0$. 

\section{Exact Solutions of Nonlocal FLE}  

Inspired by the ansatz (\ref{3}) in the case of the (local) FLE, we start from 
the same ansatz
\be\label{5.1}
u(x,t) = \phi(y) e^{i\theta(y) -i\Omega t +ikx}\,,~~ y = x- vt\,,
\ee
and substitute it in the nonlocal Eq. (\ref{1a}). On 
substituting the ansatz (\ref{5.1}) in Eq. (\ref{1a}) and assuming that 
$\theta(-y) = -\theta(y)$ (we will provide a posteri justification for this 
assumption) and separating real and 
imaginary parts we get two equations. The real part gives the equation
\be\label{5.2}
v\phi''(y) = - \sigma [k+\theta'(y)] \phi^2(y) \phi(-y) +(k\Omega-1)\phi(y) 
+v [\theta'(y)]^2 \phi(y) +(\Omega + vk) \theta'(y) \phi(y)\,.
\ee
On the other hand, the imaginary part gives the equation
\be\label{5.3}
\sigma \phi(y) \phi(-y) \phi'(y) - v\phi(y)\theta''(y) 
-[\Omega +kv +2v\theta'(y)]\phi(y) = 0\,.
\ee

We now consider two possibilities.

\subsection{Possibility I: $\phi(-y) = \phi(y)$}

In this case it is easy to see that we get exactly the same equations for both
the real and the imaginary parts (i.e. Eqs. (\ref{4}) and (\ref{5}),  
respectively) as in the local FLE case.  As in the FLE 
case, on multiplying Eq. (\ref{5.3}) by $\phi(y)$ and integrating we obtain
\be\label{5.4}
\theta'(y) = \frac{\sigma \phi^2(y)}{4v} -\frac{(\Omega+kv)}{2v} 
+ \frac{A}{v \phi^2(y)}\,,
\ee
where $A$ is a constant of integration. On substituting this expression for 
$\theta'(y)$ in Eq. (\ref{5.2}) for the real part, we find that $\phi(y)$
satisfies the $\phi^6$-like  equation as in the local case with the same expressions 
for $c_1$, $c_2$, $c_3$, $c_4$ as in the local case (see Eqs. (\ref{7}) and (\ref{8}))
\be\label{5.5}
 \phi''(y) = c_1 \phi^5(y) + c_2 \phi^3(y) + c_3 \phi(y) + c_4 \phi^{-3}(y)\,,
\ee
where
\bea\label{5.6}
&&c_1 = -\frac{3}{16v^2} < 0\,,~~ c_2 = \sigma \frac{(\omega -vk)}{2v^2}\,,
\nonumber \\
&&c_3 = -\frac{(vk-\Omega)^2 +2A\sigma+4v}{4v^2}\,,~~
c_4 = \frac{A^2}{v^2} > 0\,.
\eea
Note that while $c_1 < 0, c_4 > 0$ and $c_2, c_3$ could be
zero, negative or positive depending on the values of the parameters $v,k,
\Omega, A$. From Eq. (\ref{5.4}) we note that $\theta'(y)$ is an even function
of $y$ so that indeed $\theta(y)$ is an odd function of $y$ thereby justifying
our assumption of $\theta(-y) = -\theta(y)$.

Thus all the solutions obtained for the local case which satisfy $\phi(-y) 
= \phi(y)$ are also the solutions for the nonlocal FLE for the {\it same}
values of the parameters $c_1,c_2,c_3,c_4$. In particular, we find that out of
the 34 solutions of FLE given in Sec. III, 25 solutions satisfy 
$\phi(-y) = \phi(y)$ and hence these 25 solutions are also the solutions of the 
nonlocal FLE with exactly the same conditions on the parameters as in the local
case. In particular solutions I to V, VII, VIII, X, XII to XV, XVII to XXIV, 
XXVII, XXVIII and XXXII to XXXIV of the FLE given in Sec. III are also the 
solutions of the nonlocal FLE in case $\phi(-y) = \phi(y)$. Besides all seven  
solutions of FLE (\ref{5.5}) with $A \ne 0$ also satisfy $\phi(-y) = \phi(y)$
and hence these seven solutions are also the solutions of the nonlocal FLE with
exactly the same conditions on the parameters as in the local case.
 
\subsection{Possibility II: $\phi(-y) = -\phi(y)$}

In this case too we repeat the same steps as given by Eqs. (\ref{5.1}) to 
(\ref{5.3}) including the assumption of $\theta (-y) = \theta(y)$.
On substituting $\phi(-y) = -\phi(y)$ in Eqs. (\ref{5.2}) and (\ref{5.3}) we 
respectively obtain
\be\label{5.7}
v\phi''(y) = \sigma [k+\theta'(y)] \phi^3(y) +(k\Omega-1)\phi(y) 
+v [\theta'(y)]^2 \phi(y) +(\Omega + vk) \theta'(y) \phi(y)\,, 
\ee
\be\label{5.8}
\sigma \phi(y)^{2} \phi'(y) + v\phi(y)\theta''(y) 
+[\Omega +kv +2v\theta'(y)]\phi(y) = 0\,.
\ee
On multiplying Eq. (\ref{5.8}) by $\phi(y)$ and integrating we obtain
\be\label{5.9}
\theta'(y) = -\frac{\sigma \phi^2}{4v} -\frac{(\Omega+kv)}{2v} 
+ \frac{A}{v \phi^2}\,,
\ee
where $A$ is a constant of integration. On substituting this expression for 
$\theta'(y)$ in the Eq. (\ref{5.7}) for the real part, we find that $\phi(y)$
satisfies the $\phi^6$-like  equation which has exactly the same form as 
Eq. (\ref{7}) and with same expressions for $c_1, c_4$ while the expressions
for $c_2,c_3$ get modified, i.e. we 
again obtain
\be\label{5.10}
\phi''(y) = c_1 \phi^5 + d_2 \phi^3 + d_3 \phi + c_4 \phi^{-3}\,,
\ee
where 
\bea\label{5.11}
&&d_2 = -\sigma \frac{(\omega -vk)}{2v^2}\,,~~d_3 = -\frac{(vk-\Omega)^2 
-2A\sigma+4v}{4v^2}\,, \nonumber \\
&&c_1 = -\frac{3}{16v^2} < 0\,,~~ c_4 = \frac{A^2}{v^2} > 0\,.
\eea
Observe that $d_2 = -c_2$ 
while in $d_3$ only the $A$ dependent term changes sign compared to that in 
$c_3$. Thus, we have the solutions with the constant of integration $A = 0$, 
$d_3 = c_3$ so that only $c_2$ gets changed to $d_2$ in the case of the 
solutions satisfying $\phi(-y) = -\phi(y)$. From Eq. (\ref{5.9}) we note that 
$\theta'(y)$ is an even function of $y$ so that indeed $\theta(y)$ is an odd 
function of $y$ thereby justifying our assumption of $\theta(-y) = -\theta(y)$.

Thus, all the solutions obtained for the local case which satisfy $\phi(-y) 
= -\phi(y)$ and for which $A = 0$ are also the solutions of the nonlocal FLE 
Eqs. (\ref{5.10}) and (\ref{5.11}) except that the expression for $d_2$ is 
different from that of the local case. In particular, for $A = 0$ case, there 
are five solutions of the FLE satisfying $\phi(-y) = 
-\phi(y)$ (i.e. solutions V, IX, XI, XVI, XXXI) which therefore are also the
solutions of the nonlocal FLE as given by Eqs. (\ref{5.10}) and (\ref{5.11}) 
which we now present.

{\bf Solution I}

Eqs. (\ref{5.10}) and (\ref{5.11}) admit (hyperbolic) kink solution
\be\label{5.12}
\phi = \frac{A\sinh(\beta y)}{\sqrt{B+\cosh^2(\beta y)}}\,,~~ -1 < B < 0 \,,
\ee
provided
\be\label{5.13}
(1-|B|) c_3 = -(|B|+2)\beta^2\,,~~(1-|B|)d_2 A^2 = (2|B|+1)\beta^2\,,~~
c_1 A^4 = -3|B| \beta^2\,.
\ee
Thus for this solution $c_1$, $c_3 < 0$, $d_2 > 0$ in case $-1 < B < 0$.

{\bf Solution II}

Eqs. (\ref{5.10}) and (\ref{5.11}) also admit the kink solution 
\be\label{5.14}
\phi = \frac{Ay}{\sqrt{B+y^2}}\,,~~ B > 0\,,
\ee
with a power law tail provided
\be\label{5.15}
B c_3 = -3\,,~~B d_2 A^2 = 6\,,~~ B c_1 A^4 = -3\,.
\ee
Thus for this solution while $c_1$, $c_3 < 0$, $d_2 > 0$.

{\bf Solution III}

Eqs. (\ref{5.10}) and (\ref{5.11}) also admit the pulse solution with a power
law tail
\be\label{5.16}
\phi = \frac{Ay}{B+y^2}\,,~~ B > 0\,,
\ee
provided
\be\label{5.17}
B c_3 = -6\,,~~B d_2 A^2 = 14\,,~~ B c_1 A^4 = -8\,.
\ee
Thus for this solution while $c_1$, $c_3 < 0$, $d_2 > 0$.

{\bf Solution IV}

Eqs. (\ref{5.10})  and (\ref{5.11}) also admit a trigonometric kink solution
\be\label{5.18}
\phi = \frac{A\sin(\beta y)}{\sqrt{1+B\cos^2(\beta y)}}\,,
~~B > 1\,,
\ee
provided
\bea\label{5.19}
&&(B+1)c_3 = (2B-1)\beta^2\,,~~(B+1)d_2 A^2 = -2B(2-B)\beta^2\,, \nonumber \\
&&(B+1) c_1 A^4 = -3 B (B-1) \beta^2\,.
\eea
Thus for this solution while $c_1 < 0$, $c_3 > 0$, $d_2 = 0$ in case $B = 2$ 
while $d_2 > 0$ if $B > 2$. On the other hand, if $1 < B < 2$ then $d_2 < 0$.

{\bf Solution V}

It is easy to check that
\be\label{5.20}
\phi = \frac{A\sn(\beta y,m)}{\sqrt{1+B\cn^2(\beta y,m)}}\,,
~~B > \frac{1}{1-m}\,,
\ee
is an exact periodic kink solution of Eqs. (\ref{5.10}) and (\ref{5.11}) 
provided
\bea\label{5.21}
&&(B+1) c_3 = [(2-m)B-(1+m)]\beta^2\,,~~(B+1)d_2 A^2 
= -2[(2B-m) -(1-m)B^2]\beta^2\,, \nonumber \\
&&c_1 A^4 = -3B[(1-m)B-1] \beta^2\,.
\eea 
For this solution while $c_1 < 0$, the values of $d_2, c_3$ depend on
the values of $m$ and $B$.

Before completing this section it is worth pointing out that there are four  
solutions of the FLE (i.e. Solutions XXV, XXVI, XXIX, XXX) which do not satisfy
$\phi(-y) = \pm \phi(y)$ and hence are {\it not} the solutions of the nonlocal 
FLE even though they are the solutions of the local FLE.

\section{One-Parameter Family of FLEs} 

Motivated by the generalization of NLS to arbitrary power, we generalize
the FLE to a one-parameter family of FLEs characterized by the
parameter $n$ where $n$ is a positive integer or half-integer, and show that 
one can reduce the problem to the solutions of the 
$\phi^{4n+2}$-$\phi^{2n+2}$-$\phi^2$ field theory. 

Let us consider the following one-parameter family of FLE
\be\label{6.1}
u_{xt}(x,t) + u(x,t) -i\sigma |u|^{2n} u_x(x,t) = 0\,,
\ee
where $n  = 1/2, 1, 3/2, 2,...$ is either a positive half-integer or an 
integer. We start from the ansatz
\be\label{6.2}
u(x,t) = \phi(y) e^{i\theta(y) -i\Omega t +ikx}\,,~~ y = x- vt\,,
\ee
and substitute it in Eq. (\ref{6.1}) and separate the real and imaginary
parts. While real part gives the equation
\be\label{6.3}
v\phi''(y) = - \sigma [k+\theta'(y)] \phi^{2n +1} (y) +(k\Omega-1)\phi(y) 
+v [\theta'(y)]^2 \phi(y) +(\Omega + vk) \theta'(y) \phi(y)\,,
\ee
the imaginary part leads to the equation
\be\label{6.4}
\sigma \phi(y)^{2n} \phi'(y) - v\phi(y)\theta''(y) 
-[\Omega +kv +2v\theta'(y)]\phi(y) = 0\,.
\ee
On multiplying Eq. (\ref{6.4}) by $\phi(y)$ and integrating we obtain
\be\label{6.5}
\theta'(y) = \frac{\sigma \phi^{2n}(y)}{4v} -\frac{(\Omega+kv)}{2v} 
+ \frac{A}{v \phi^2(y)}\,,
\ee
where $A$ is a constant of integration. On substituting this expression for 
$\theta'(y)$ in Eq. (\ref{6.3}) for the real part, we find that $\phi(y)$
satisfies the equation
\be\label{6.6}
\phi''(y) = c_1 \phi^{4n+1}(y) + c_2 \phi^{2n+1}(y) + c_3 \phi(y) 
+ c_4 \phi^{-3}(y) + c_5 \phi^{2n -1}\,,
\ee
where
\bea\label{6.7}
&&c_1 = -\frac{(2n+1}{4 v^2 (n+1)^2} < 0\,,~~ c_2 = 
\sigma \frac{(\omega -vk)}{2v^2}\,,
\nonumber \\
&&c_3 = -\frac{(vk-\Omega)^2+4v}{4v^2}\,,~~c_4 = \frac{A^2}{v^2} > 0\,,
c_5 = -\frac{n\sigma A}{v^2 (n+1)}\,.  
\eea
Note that while $c_1 < 0$, $c_4 > 0$ and $c_2$, $c_3$, $c_5$ could be
zero, negative or positive depending on the values of the parameters $v,k,
\Omega, A$. 

Several comments are in order here.

\begin{enumerate}

\item In the special case of $n = 1$, as expected Eq. (\ref{6.6}) goes over
to the $\phi^6$-like  equation for FLE and the $c_5$ term becomes a part of 
the $c_3$ term.
		
\item In the special case of the constant of integration $A = 0$, $c_5$ and 
$c_4$ terms disappear.

\item In case $n$ is an integer, Eq. (\ref{6.6}) is the field equation 
for the $\phi^{4n+2}$-$\phi^{2n+2}$-$\phi^2$ field theory.

\item On the other hand, in the special case of $n = 1/2$,  one gets the field 
equation for the asymmetric  $\phi^4$ field theory.

\end{enumerate}

\subsection{Exact Solutions of Eq. (\ref{6.6}) in Case $A = 0$}

We now present four solutions of Eqs. (\ref{6.6}) and (\ref{6.7}) 
in case the constant of integration is zero. Thus
we want the exact solutions of the equation
\be\label{6.8}
\phi''(y) = c_1 \phi^{4n+1}(y) + c_2 \phi^{2n+1}(y) + c_3 \phi(y)\,,
\ee
where
\bea\label{6.9}
&&c_1 = -\frac{(2n+1}{4 v^2 (n+1)^2} < 0\,,~~ c_2 = 
\sigma \frac{(\Omega -vk)}{2v^2}\,,
\nonumber \\
&&c_3 = -\frac{(vk-\Omega)^2+4v}{4v^2}\,,
\eea

{\bf Solution I}

It is easily checked that
\be\label{6.10}
\phi(y) = A \sech^{1/2n}(\beta y)\,,
\ee
is an exact solution of Eqs. (\ref{6.6}) and (\ref{6.7}) provided
\be\label{6.11}
c_2 = 0\,,~~c_3 = \frac{\beta^2}{4n^2}\,,~~c_1 A^{4n} = -\frac{(2n+1) \beta^2}
{4n^2}\,.
\ee

{\bf Solution II}

Another solution to Eqs. (\ref{6.6}) and (\ref{6.7}) is
\be\label{6.12}
\phi(y) = \frac{A}{[B+\cosh(\beta y)]^{1/2n}}\,,~~ 0 \le B < 1\,,
\ee
provided
\be\label{6.13}
c_3 = \frac{\beta^2}{4n^2}\,,~~c_2 A^{2n} = -\frac{(n+1)B \beta^2}{2n^2}\,,~~
c_1 A^{4n} = -\frac{(1-B^2)(2n+1)\beta^2}{4n^2} < 0\,.
\ee

It is worth noticing that in the limit $B = 0$, the solution II goes over to
the solution I and the conditions (\ref{6.13}) go over to the conditions
(\ref{6.11}).

{\bf Solution III}

Interestingly, Eqs. (\ref{6.6}) and (\ref{6.7}) also admit a trigonometric 
solution
\be\label{6.14}
\phi(y) = \frac{A}{[B+\cos(\beta y)]^{1/2n}}\,,~~ B > 1\,,
\ee
provided
\be\label{6.15}
c_3 = -\frac{\beta^2}{4n^2}\,,~~c_2 A^{2n} = \frac{(n+1)B \beta^2}{2n^2}\,,~~
c_1 A^{4n} = -\frac{(B^2-1)(2n+1)\beta^2}{4n^2} < 0\,.
\ee

{\bf Solution IV}

Eqs. (\ref{6.6}) and (\ref{6.7}) also admit a pulse solution with
a power law tail
\be\label{6.16}
\phi(y) = \frac{A}{(B+ y^2)^{1/2n}}\,,~~ B > 0\,,
\ee
provided
\be\label{6.17}
c_3 = 0\,,~~c_2 A^{2n} = \frac{(n+1)}{n^2}\,,~~
c_1 A^{4n} = -\frac{(Bn+n+1)}{n^2} < 0\,.
\ee

\section{Conclusions and Open Problems}

In this paper we have obtained a variety of 34 solutions of the integrable FLE 
in case the constant of integration is zero and seven solutions in case the 
constant of integration is nonzero. While some of the solutions are already 
known, we believe that many of our solutions (especially the periodic ones) and 
those when the constant of integration is nonzero are new. Besides we have also
considered a nonlocal integrable FLE and shown that it admits 37 out of the 
41 solutions admitted by FLE, most under the same conditions while five of them
under different conditions. Finally, we have generalized FLE to a one-parameter
family of FLEs and obtained some of its solutions and in one special case have
obtained its large number of solutions. This paper raises several open 
questions some of which are the following.

\begin{enumerate}

\item While we have obtained a large number of solutions of FLE and nonlocal 
FLE it is not obvious if we have exhausted all possible solutions. It would
be interesting to find out more solutions of these nonlinear equations.

\item In recent years FLE has found application in optics. It is then worth
finding out if some of the solutions that we have obtained have relevance 
in the optical and other physical contexts. A related question is about the 
possible application of the generalized FLE that we have obtained in case 
$n = 1/2$ which is related to the quadratic NLS. 

\item Another important issue is about the stability of the solutions obtained
in this paper.

\item Recently a novel connection has been shown between FLE and a spin 
model \cite{avadh}. It is then natural to enquire if there is a spin model 
related to the integrable nonlocal FLE? Similarly are there spin models which 
are related to the one-parameter family FLE that we have proposed in this 
paper? 

\end{enumerate}

We hope to address some of these issues in the near future. \\ 

\noindent{\bf Acknowledgment}

One of us (AK) is grateful to Indian National Science Academy (INSA) for the
award of INSA Honorary Scientist Position at Savitribai Phule Pune University.
The work at NANL was carried out under the auspices of the US Department 
of Energy NNSA under contract No. 89233218CNA000001.

\section{Appendix A: Exact Solutions in Case $n = 1/2$ and $A = 0$}

We now present 24 solutions of Eqs. (\ref{6.6}) and (\ref{6.7}) in case 
$n = 1/2$ while the constant of integration $A$ is zero. Thus
we want exact solutions of the equation \cite{ks2,ks3,ks4,ks5}
\be\label{A1}
\phi''(y) = c_1 \phi^{3}(y) + c_2 \phi^{2}(y) + c_3 \phi(y)\,,
\ee
where
\be\label{A2}
c_1 = -\frac{2}{9 v^2} < 0\,,~~ c_2 = \sigma \frac{(\omega -vk)}{2v^2}\,,
c_3 = -\frac{(vk-\Omega)^2+4v}{4v^2}\,.
\ee
Out of these 24 solutions, there are five hyperbolic, two trigonometric and 15 
periodic solutions in terms of Jacobi elliptic functions and two algebraic 
solutions with a power law tail.

\subsection{Hyperbolic Solutions}

We now present five hyperbolic pulse solutions.

{\bf Solution I}

It is easy to check that Eqs. (\ref{A1}) and (\ref{A2}) admit the hyperbolic
pulse solution
\be\label{A3}
\phi = A + B \sech(\beta y)\,,
\ee
provided
\be\label{A4}
B^2 = A^2\,,~~c_1 A^2 = -\beta^2\,,~~c_3 = 2 c_1 A^2\,,~~c_2  = 3 c_1 A\,. 
\ee
Thus this solution is only valid if $c_1$, $c_2$, $c_3 < 0$.

{\bf Solution II}

Another hyperbolic pulse solution to EPS. (\ref{A1}) and (\ref{A2}) is
\be\label{A5}
\phi = B + \frac{A}{D+\sech(\beta x)}\,,~~D > 1\,,
\ee
provided 
\be\label{A6} 
(c_3-c_2 B+ c_1 B^2)B = 2AD \beta^2\,,
\ee
\be\label{A7}
(c_3-2c_2 B+3 c_1 B^2) = (1-6D^2) \beta^2\,,
\ee
\be\label{A8}
(3 c_1 B - c_2)A = 3D(2D^2-1) \beta^2\,, 
\ee
\be\label{A9}
c_1 A^2 = -2D^2 (D^2-1)\beta^2 < 0\,. 
\ee
One can solve Eqs. (\ref{A6}) to (\ref{A9}) 
and obtain a cubic equation for the ratio $B/A = y$ in terms of $D$ 
\be\label{A10}
2D^2 (D^2-1)y^3 +3D(2D^2-1) y^2 + (6D^2-1) y +2D = 0\,.
\ee
This cubic equation is easily solved and the three roots are
\be\label{A11}
y = -\frac{1}{D}\,,~~y = y_{\pm} = -\frac{(4D^2-1) \pm \sqrt{8D^2+1}}
{4D(D^2 -1)}\,.
\ee

{\bf Case I: $y = -\frac{1}{D}$}

In this case one can show that 
\be\label{A12}
 c_1 A^2 = -2D^2(D^2-1)\beta^2\,,~~c_2 A = -3D\beta^2\,,~~c_3 = \beta^2\,.
\ee
Thus for this solution $c_1, c_2 < 0, c_3 > 0$.

{\bf Case II: $y = y_{\pm}$}

In both cases we have
\be\label{A13}
c_2 A = 3 c_1  A^2 y_{\pm} -3D(2D^2-1)\beta^2\,,~~ c_3 = 3 c_1 A^2 y_{\pm}^{2} 
-6D(2D^2-1) y_{\pm}\beta^2 -(6D^2-1)\beta^2\,,
\ee
so that in these cases $c_1, c_2, c_3 < 0$.

{\bf Solution III}

Eqs. (\ref{A1}) and (\ref{A2}) also admit another hyperbolic pulse 
solution
\be\label{A14}
\phi = \frac{A}{B+\cosh^2(\beta y)}\,,~~ -1 <B < 0\,,
\ee
provided 
\be\label{A15} 
c_3 = 4\beta^2\,,~~ c_2 A = 6(1-2|B|)\beta^2\,,~~ c_1 A^2 
= -8|B|(1-|B|) \beta^2\,.
\ee
Thus for this solution $c_1 < 0, c_3 > 0$ while $c_2 > (<) = 0$ depending on
if $B < (>) = 1/2$.

{\bf Solution IV}

Eqs. (\ref{A1}) and (\ref{A2}) also admit another hyperbolic pulse 
solution
\be\label{A16}
\phi = D - \frac{A}{B+\cosh^2(\beta x)}\,,~~ -1 < B < 0\,,
\ee
provided 
\be\label{A17} 
c_2 D = c_3 + c_1 D^2\,,
\ee
\be\label{A18}
c_1 A^2 = -8|B|(1-|B|)\beta^2\,,
\ee
\be\label{A19}
4\beta^2 = c_1 D^2 -c_3\,,
\ee
\be\label{A20}
(3c_1 D +c_2)A = -6(1-2|B|) \beta^2\,.
\ee
One can solve Eqs. (\ref{A17}) to (\ref{A20}) and obtain a quadratic 
equation for the ratio $D/A = y$ in terms of $|B|$. We obtain
\be\label{A21}
4|B|(1-|B|)y^2 -3(2|B|-1)y-2 = 0\,,
\ee
which on solving gives
\be\label{A22}
y = \frac{D}{A} = \frac{3(2|B|-1) \pm \sqrt{4B^2-4|B|+9}}{8|B|(1-|B|)}\,.
\ee
One can now compute $c_2, c_3$.

{\bf Solution V}

Another solution to Eqs. (\ref{A1}) and (\ref{A2}) is
\be\label{A23}
\phi = B + \frac{A\sech(\beta y)}{D+\tanh(\beta y)}\,,~~D > 1\,,
\ee
provided 
\be\label{A24}
c_1, c_2, c_3 < 0\,,~~c_3 = 2 c_1 B^2\,,~~c_2 = 3 c_1 B\,,
~~|c_1| B^2 = \beta^2\,,~~|c_1| A^2 = 2(D^2-1)\beta^2\,.
\ee
Thus for this solution $c_1, c_2, c_3 < 0$. 

\subsection{Trigonometric Solutions}

We now present two trigonometric solutions.

{\bf Solution VI}

Eqs. (\ref{A18}) and (\ref{A19}) also admit a trigonometric pulse solution
\be\label{A25}
\phi = \frac{A}{D+\cos(\beta y)}\,,~~B > 1\,,
\ee
provided 
\be\label{A26}
c_1, c_2, c_3 < 0\,,~~|c_3| = \beta^2\,,~~|c_2|A = 3D \beta^2\,,~~|c_1|A^2 
= 2(D^2-1)\beta^2\,.
\ee
Thus this solution is only valid if $c_1, c_2, c_3 < 0$.

{\bf Solution VII}

Yet another trigonometric solution to Eqs. (\ref{A18}) and (\ref{A19}) is
\be\label{A27}
\phi = \frac{A+B\cos(\beta y)}{D+\cos(\beta y)}\,,~~D > 1\,,
\ee
provided
\bea\label{A28}
&&(B D-A) c_1 B^2 = -(A+2BD)\beta^2\,,~~(BD-A) c_2 B = 3((A+BD)\beta^2\,,
\nonumber \\
&&(BD-A) c_3 = -(2A+BD)\beta^2\,.
\eea
Since we want $c_1 < 0$ hence we demand that $A, B, D > 0$ and further 
$BD > A > 0$.

\subsection{Algebraic Solutions With Power Law Tail}

We now present two algebraic solutions of Eqs. (\ref{A1}) and (\ref{A2}). 

{\bf Solution VIII}

It is easily checked that Eqs. (\ref{A18}) and (\ref{A19}) admit the solution
\be\label{A29}
\phi = \frac{A}{B +y^2}\,,~~ B > 0\,,
\ee
provided
\be\label{A30}
c_3 = 0\,,~~ c_1, c_2 < 0\,,~~|c_2|A = 6\,,~~|c_1|A^2 = 8B\,.
\ee
Thus this solution is only valid if $c_3 = 0$ and $c_1, c_2 < 0$.

{\bf Solution IX}

It is easily checked that Eqs. (\ref{A1}) and (\ref{A2}) also admit another 
algebraic solution with a power law tail
\be\label{A31}
\phi = \frac{A(y^2+B)}{(D +y^2)}\,,~~ B, D > 0\,,~~ A < 0 \,,
\ee
provided
\be\label{A32}
c_3 = c_1 A^2\,,~~ c_2 = - 2c_1 A\,,~~|c_1|A^2 = \frac{-9}{D}\,.
\ee
Thus this solution is only valid if $c_1, c_3 < 0$ and $c_2 > 0$.

\subsection{Periodic Solutions in Terms of Jacobi Elliptic Functions}

We will now present nine  periodic solutions in terms of Jacobi elliptic functions
of Eqs. (\ref{A1}) and (\ref{A2}).

{\bf Solution X}

It is easily checked that Eqs. (\ref{A1}) and (\ref{A2}) admit a 
periodic pulse solution
\be\label{A33}
\phi = A + B \dn(\beta y,m)\,,~~ 0 < m \le 1\,,
\ee
provided
\be\label{A34}
c_1, c_2, c_3 < 0\,,~~ c_3 = 2 c_2 A^2\,,~~c_2 = 3c_1 A\,,~~|c_1| A^2 
= (2-m)\beta^2\,,~~|c_1| B^2 = 2\beta^2\,.
\ee
Thus this solution is only valid if $c_1, c_2, c_3 < 0$.

{\bf Solution XI}

It is easily checked that Eqs. (\ref{A1}) and (\ref{A2}) admit another
periodic pulse solution
\be\label{A35}
\phi = A + B \cn(\beta y,m)\,,~~ 1/2 < m \le 1\,,
\ee
provided
\be\label{A36}
c_1, c_2, c_3 < 0\,,~~ c_3 = 2 c_2 A^2\,,~~c_2 = 3c_1 A\,,~~|c_1| A^2 
= (2m-1)\beta^2\,,~~|c_1| B^2 = 2\beta^2\,.
\ee
Thus this solution is only valid if $c_1, c_2, c_3 < 0$.

{\bf Solution XII}

Remarkably, Eqs. (\ref{A1}) and (\ref{A2}) also admit  the 
superposition of solutions X and XI as a solution
\be\label{A37}
\phi = A + B[\dn(\beta y,m)+\sqrt{m} \cn(\beta y,m)]\,,~~0 < m < 1\,,
\ee
provided
\be\label{A38}
c_1, c_2, c_3 < 0\,,~~c_3 = 2 c_1 A^2\,,~~c_2 = 3c_1 A\,,~~2|c_1| B^2 
= \beta^2\,,~~ |c_1| A^2 = \beta^2\,.
\ee
Thus this solution is only valid if $c_1, c_2, c_3 < 0$.

{\bf Solution XIII}

Eqs. (\ref{A1}) and (\ref{A2}) also admit another periodic pulse solution 
\be\label{A39}
\phi = A + \frac{B\sqrt{1-m}}{\dn(\beta y,m)}\,, ~~ 0 < m < 1\,,
\ee
provided
\be\label{A40}
c_1, c_2, c_3 < 0\,,~~ c_3 = 2 c_2 A^2\,,~~c_2 = 3c_1 A\,,~~|c_1| A^2 
= (2-m)\beta^2\,,~~|c_1| B^2 = 2\beta^2\,.
\ee
Thus this solution is only valid if $c_1, c_2, c_3 < 0$.

{\bf Solution XIV}

Remarkably, not only $\dn(y,m)$ and $1/\dn(y,m)$ but even their superposition is an 
exact solution of Eqs. (\ref{A1}) and (\ref{A2}). In particular, it is easily
checked that
\be\label{A41}
\phi  = D + A\dn(\beta y,m) +\frac{B\sqrt{1-m}}{\dn(\beta y,m)}\,,
\ee
is also an exact solution of Eqs. (\ref{A1}) and (\ref{A2}) provided
\bea\label{A42}
&&B = \pm A\,,~~ c_3 = 2 D^2 c_1\,,~~ c_2 = -3 D c_1\,,~~|c_1| A^2 = 2\beta^2\,,
~~ m > 4(3\sqrt{2}-4)\,, \nonumber \\
&&|c_1| D^2 = [(2-m) \pm 6\sqrt{1-m}]\beta^2\,.
\eea
Thus this solution exists only if $m$ is greater than $4(3\sqrt{2}-4)$ which 
is approximately $0.92$.

{\bf Solution XV}

Eqs. (\ref{A1}) and (\ref{A2}) also admit another periodic pulse solution 
\be\label{A43}
\phi = A + \frac{B\sqrt{m(1-m)}\sn(\beta y,m)}{\dn(\beta y,m)}\,, ~~ 
1/2 < m < 1\,,
\ee
provided
\be\label{A44}
c_1, c_2, c_3 < 0\,,~~ c_3 = 2 c_2 A^2\,,~~c_2 = 3c_1 A\,,~~|c_1| A^2 
= (2m-1)\beta^2\,,~~|c_1| B^2 = 2\beta^2\,.
\ee
Thus this solution is only valid if $c_1, c_2, c_3 < 0$.

{\bf Solution XVI}

Remarkably, not only $1/\dn(y,m)$ and $sn/\dn(y,m)$ but even their superposition is an 
exact solution of Eqs. (\ref{A1}) and (\ref{A2}). In particular, it is easily
checked that
\be\label{A45}
\phi  = D + A\dn(\beta y,m) +\frac{B\sqrt{m(1-m)}\sn(\beta y,m)}
{\dn(\beta y,m)}\,,
\ee
is also an exact solution of Eqs. (\ref{A1}) and (\ref{A2}) provided
\be\label{A45a}
B = \pm A\,,~~ c_3 = 2 D^2 c_1\,,~~ c_2 = -3 D c_1\,,~~ 2|c_1| A^2 
= \beta^2\,,~~ |c_1| D^2 = \frac{(1+m)\beta^2}{2}\,.
\ee
Thus this solution is valid only if $c_1, c_3 < 0\,,~~c_2 > 0$.

{\bf Solution XVII}

Eqs. (\ref{A1}) and (\ref{A2}) also admit the periodic solution
\be\label{A46}
\phi = B + \frac{A\dn(\beta y,m)}{D+\sn(\beta y,m)}\,,~~ 0 < m < 1\,,
\ee
provided 
\bea\label{A47}
&&c_1, c_2, c_3 < 0\,,~~c_3 = 2 c_1 B^2\,,~~c_2 = 3 c_1 B\,, \nonumber \\
&&|c_1| B^2 = \frac{(1+m)\beta^2}{2}\,, ~~ \sqrt{m} D =1\,,~~
|c_1| A^2 = \frac{(1-m)\beta^2}{2m}\,.
\eea
Thus this solution is only valid if $c_1, c_2, c_3 < 0$.

{\bf Solution XVIII}

Eqs. (\ref{A1}) and (\ref{A2}) also admit the periodic superposed 
solution
\be\label{A48}
\phi = F + \frac{A[\dn(\beta y,m)+B\sqrt{m}\cn(\beta y,m)]}
{D+\sn(\beta y,m)}\,,~~ D > 1/\sqrt{m}\,,
\ee
provided 
\bea\label{A49}
&&c_1, c_2, c_3 < 0\,,~~c_3 = 2 c_1 B^2\,,~~c_2 = 3 c_1 B\,,~~
|c_1| F^2 = -\frac{(1+m)\beta^2}{2}\,, \nonumber \\
&&|c_1| A^2 = \frac{(D^2-1)\beta^2}{2}\,, ~~ 0 < m < 1\,,~~
|c_1| B^2 = \frac{(m D^2 -1)\beta^2}{2m}\,.
\eea
Thus this solution is only valid if $c_1, c_2, c_3 < 0$.

{\bf Solution XIX}

Eqs. (\ref{A1}) and (\ref{A2}) also admit a periodic kink solution
\be\label{A50}
\phi = D - \frac{A\sqrt{m}\sn(\beta y,m)}{1+B\cn^2(\beta y,m)}\,,~~B > 0\,,
~~ m > 1/36\,,
\ee
provided 
\bea\label{A51}
&&c_1, c_2, c_3 < 0\,,~~c_3 = 2 c_1 D^2\,,~~c_2 = 3c_1 D\,,~~|c_1| D^2 
= (6\sqrt{m}-1-m)\beta^2 > 0\,,
\nonumber \\
&&|c_1| A^2 = 8m\beta^2\,,~~B = \frac{\sqrt{m}}{(1-\sqrt{m})}\,.
\eea
Thus this solution is only valid if $c_1, c_2, c_3 < 0$ and $6\sqrt{m} > 1+m$
which implies that $m > 1/36$. 

{\bf Solution XX}

Eqs. (\ref{A1}) and (\ref{A2}) also admit a periodic pulse solution
\be\label{A52}
\phi = D - \frac{A\dn(\beta y,m)}{1+B\cn^2(\beta y,m)}\,,~~B > 0\,,~~
m \ge 1/3\,,
\ee
provided 
\bea\label{A53}
&&c_1, c_2, c_3 < 0\,,~~c_3 = 2 c_1 D^2\,,~~c_3= 3 c_1 D\,,~~
|c_1| D^2 = (2-m+6\sqrt{m})\beta^2\,,  \nonumber \\
&&|c_1| A^2 =  \frac{8\beta^2}{\sqrt{1-m}}\,,~~
B = \frac{1-\sqrt{1-m}}{(\sqrt{1-m})}\,.
\eea
Thus this solution is only valid if $c_1$, $c_2$, $c_3  < 0$.
Further one finds that in this case $m \ge 1/3$.

{\bf Solution XXI}

Eqs. (\ref{A1}) and (\ref{A2}) also admit a periodic kink solution
\be\label{A54}
\phi = D - \frac{A\cn(\beta y,m)\sn(\beta y,m)}{1+B\cn^2(\beta y,m)}\,,
~~B > 0\,,~~m > 4(3\sqrt{2}-4)\,,
\ee
provided 
\bea\label{A55}
&&c_1, c_2, c_3 < 0\,,~~c_3 = 2 c_1 D^2\,,~~c_2 = 3c_1 D\,,~~|c_1| D^2 
= (2-m-6\sqrt{1-m})\beta^2\,,
\nonumber \\
&&|c_1| A^2 = \frac{8(1-\sqrt{1-m})^2 \beta^2}{\sqrt{1-m}}\,,
~~B = \frac{1-\sqrt{1-m}}{\sqrt{1-m}}\,.
\eea
Thus this solution is only valid if $c_1, c_2, c_3 < 0$ and 
$2-m > 6\sqrt{1-m}$ which implies that $m > 4(3\sqrt{2}-4)$, i.e.
approximately $m > 0.92$.

{\bf Solution XXII}

Eqs. (\ref{A1}) and (\ref{A2}) also admit a periodic pulse solution
\be\label{A56}
\phi = B + \frac{A}{D+\dn(\beta y,m)}\,,
\ee
provided 
\be\label{A57} 
(c_3+c_2 B+ c_1 B^2)B = 2 A D \beta^2\,,
\ee
\be\label{A58}
(c_3 +2c_2 B +3 c_1 B^2) = (2-m-6 D^2) \beta^2\,,
\ee
\be\label{A59}
(3 c_1 B +c_2)A = 3D[2D^2-(2-m)] \beta^2\,,
\ee
\be\label{A60}
c_1 A^2 = -2(D^2-1)[D^2-(1-m)]\beta^2\,.
\ee
Thus for this solution either $D^2 < 1-m$ or $D^2 > 1$ so that $c_1 < 0$.
One can in fact solve Eqs. (\ref{A57}) to (\ref{A60}) 
and obtain a cubic equation for the ratio $A/B = y$ in terms of $D$ 
and $m$
\be\label{A61}
2D y^3 +[6D^2-(2-m)]y^2 +3D[2D^2-(2-m)]+ 2(D^2-1)[D^2-(1-m)] = 0\,.
\ee
One can simplify the cubic equation by substituting $y = z-D$ in which case
the cubic equation in $z$ has a simpler form
\be\label{A62}
2D z^3 -(2-m)z^2 -(2-m)Dz +2(1-m) = 0\,.
\ee
Unfortunately we are unable to solve this cubic equation analytically and
obtain $y$ in terms of $D$ and $m$. 

{\bf Solution XXIII}

Eqs. (\ref{A1}) and (\ref{A2}) also admit a eriodic pulse solution
\be\label{A63}
\phi = B + \frac{A}{D+\cn(\beta y,m)}\,,~~ D > 1\,,
\ee
provided 
\be\label{A64} 
(c_3+c_2 B+ c_1 B^2)B = 2 A m D \beta^2\,,
\ee
\be\label{A65}
(c_3 +2c_2 B +3 c_1 B^2) = (2m-1-6m D^2) \beta^2\,,
\ee
\be\label{A66}
(3 c_1 B +c_2)A = 3D[2m D^2-(2m-1)] \beta^2\,, 
\ee
\be\label{A67}
c_1 A^2 = -2(D^2-1)[m D^2 +(1-m)]\beta^2\,. 
\ee
Thus for this solution $D^2 > 1$ so that $c_1 < 0$.
One can solve Eqs. (\ref{A82}) to (\ref{A85}) 
and obtain a cubic equation for the ratio $A/B = y$ in terms of $D$ 
and $m$
\be\label{A68}
2m D y^3 +[6m D^2-(2m-1)]y^2 +3D[2m D^2-(2m-1)]+ 2(D^2-1)[mD^2+(1-m)] = 0\,.
\ee
One can simplify the cubic equation by substituting $y = z-D$ in which case
the cubic equation in $z$ has a simpler form
\be\label{A69}
2m D z^3 -(2m-1)z^2 -(2m-1)Dz -2(1-m) = 0\,.
\ee
Unfortunately we are unable to solve this cubic equation analytically and
obtain $z$ in terms of $D$ and $m$ except when $m = 1/2$. In that case the
cubic Eq. (\ref{A69}) is easily solved yielding $z = D^{-1/3}$ and hence
$A/B = y = z -D = D^{-1/3} - D < 0$.

{\bf Solution XXIV}

Eqs. (\ref{A1}) and (\ref{A2}) also admit periodic kink solution
\be\label{A70}
\phi = B + \frac{A}{D+\sn(\beta y,m)}\,,~~D > 1\,,~~ 1 < D^2 < 1/m\,,
\ee
provided 
\be\label{A71} 
(c_3 +c_2 B +c_1 B^2)B = -2m AD \beta^2\,,
\ee
\be\label{A72}
(c_3+ 2c_2 B+3 c_1 B^2)B = [6m D^2 -(1+m)] \beta^2\,,
\ee
\be\label{A73}
(3 c_1 +c_2)A = -3D[2m D^2-(1+m)] \beta^2\,, 
\ee
\be\label{A74}
|c_1| A^2 = 2(D^2-1)(1-m D^2)\beta^2\,. 
\ee
Thus for this solution $c_1 < 0$ if $1 < D^2 < 1/m$. 

One can in fact solve Eqs. (\ref{A71}) to (\ref{A74}) 
and obtain a cubic equation for the ratio $A/B = y$ in terms of $D$ 
and $m$
\be\label{A75}
2m D y^3 +[6m D^2 -(1+m)]y^2 +3D[2m D^2 -(1+m)]y + 2(D^2-1)(m D^2-1) = 0\,.
\ee
One can simplify the cubic equation by substituting $y = z-D$ in which case
the cubic equation in $z$ has a simpler form
\be\label{A76}
2m D z^3 -(1+m)z^2 -(1+m)Dz -2 = 0\,.
\ee
Unfortunately we are unable to solve this cubic equation analytically and
obtain $y$ in terms of $D$ and $m$ except in the special case of 
$D^2 = 2, m = 1/4$.

{\bf Special Case: $D^2 = 2, m = 1/4$}

In this case the three roots are 
\be\label{A77}
y = -\frac{1}{2\sqrt{2}}\,,~~y_{\pm} = \frac{-1 \pm\sqrt{17}}{2\sqrt{2}}\,.
\ee
It is then straightforward to compute corresponding $c_1, c_2, c_3$.

\subsection{Appendix AI: Exact Solutions in Case $n = 1/2, c_2 = 0$ and 
$A = 0$}

It turns out that in the special case of $c_2 = 0$, i.e. when $\Omega = vk$
one has several simpler solutions which we now present one by one. In 
particular, we now present 15 solutions of Eqs. (\ref{A1}) and (\ref{A2}) in 
case $n = 1/2, c_2 = 0$ while the constant of integration is zero. 
Thus we want exact solutions of the equation \cite{aubry,fred,ks2,ks3,ks4,ks5}
\be\label{A78}
\phi''(y) = c_1 \phi^{3}(y) + c_3 \phi(y)\,,
\ee
where
\be\label{A79}
c_1 = -\frac{(2}{9 v^2} < 0\,,~~ c_3 = -\frac{1}{v}\,.
\ee
Notice that while $c_1 < 0$, $c_3$ can be positive or negative depending on
if $v > (<)$ 0.

Out of these 15 solutions, there are two hyperbolic, and 13 
periodic solutions in terms of Jacobi elliptic functions.

\subsubsection{Hyperbolic Solutions}

We now present two hyperbolic pulse solutions of Eqs. (\ref{A78}) and 
(\ref{A79}).

{\bf Solution XXV}

It is easy to check that Eqs. (\ref{A78}) and (\ref{A79}) admit the hyperbolic
pulse solution
\be\label{A80}
\phi =  A \sech(\beta y)\,,
\ee
provided
\be\label{A81}
c_1 A^2 = -\beta^2\,,~~c_3 = \beta^2\,.
\ee
Thus this solution is only valid if $c_1 < 0, c_3 > 0$.

{\bf Solution XXVI}

It is easy to check that Eqs. (\ref{A78}) and (\ref{A79}) admit the hyperbolic
solution
\be\label{A82}
\phi =  \frac{A\sech(\beta y)}{D+\tanh(\beta y)}\,,~~ D > 1\,,
\ee
provided
\be\label{A83}
c_1 A^2 = -2(D^2-1) \beta^2\,,~~c_3 = \beta^2\,.
\ee
Thus this solution is only valid if $c_1 < 0, c_3 > 0$.

\subsubsection{Periodic Solutions in Terms of Jacobi Elliptic Functions}

We will now present nine periodic solutions in terms of Jacobi elliptic functions
of Eqs. (\ref{A78}) and (\ref{A79}).

{\bf Solution XXVII}

It is easily checked that Eqs. (\ref{A78}) and (\ref{A79}) admit a 
periodic pulse solution
\be\label{A84}
\phi = A \dn(\beta y,m)\,,
\ee
provided
\be\label{A85}
c_1 A^2 = -2\beta^2\,,~~c_3 = (2-m)\beta^2\,.
\ee
Thus this solution is only valid if $c_1 < 0, c_3 > 0$.

{\bf Solution XXVIII}

It is easily checked that Eqs. (\ref{A78}) and (\ref{A79}) admit another
periodic pulse solution
\be\label{A86}
\phi = A \cn(\beta y,m)\,,
\ee
provided
\be\label{A87}
c_1 A^2 = -2\beta^2\,,~~c_3 = (2m-1)\beta^2\,.
\ee
Thus this solution is only valid if $c_1 < 0$ while $c_3 > (<)$ 0 depending on
if $m > (<)$ 1/2.

{\bf Solution XXIX}

Remarkably Eqs. (\ref{A78}) and (\ref{A79}) also admit a 
superposition of solutions II and III as a solution
\be\label{A88}
\phi = A \dn(\beta y,m) + B\sqrt{m} \cn(\beta y,m)]\,,~~0 < m < 1\,,
\ee
provided
\be\label{A89}
B^2 = A^2\,,~~2c_1 A^2 = -\beta^2\,,~~ c_3 = -\frac{(1+m)\beta^2}{2}\,.
\ee
Thus this solution is only valid if $c_1, c_3 < 0$.

{\bf Solution XXX}

Eqs. (\ref{A78}) and (\ref{A79}) also admit another periodic pulse solution 
\be\label{A90}
\phi = \frac{A\sqrt{1-m}}{\dn(\beta y,m)}\,, ~~ 0 < m < 1\,,
\ee
provided
\be\label{A91}
c_1 A^2 = -2\beta^2\,,~~c_3 = (2-m)\beta^2\,.
\ee
Thus this solution is only valid if $c_1 < 0, c_3 > 0$.

{\bf Solution XXXI}

Remarkably, not only $\dn(y,m)$ and $1/\dn(y,m)$ but even their superposition is an 
exact solution of Eqs. (\ref{A78}) and (\ref{A79}). In particular, it is easily
checked that
\be\label{A92}
\phi  = A\dn(\beta y,m) +\frac{B\sqrt{1-m}}{\dn(\beta y,m)}\,,
\ee
is also an exact solution of Eqs. (\ref{A78}) and (\ref{A79}) provided
\be\label{A93}
B^2 = A^2\,,~~ c_1 A^2 = -2\beta^2\,,~~c_3 = [2-m \pm 6\sqrt{1-m}]\beta^2\,.
\ee
Thus this solution exists only if $c_1 < 0$ while $c_3 > (<)$ 0 depending on
if $2-m > (<)$  $6\sqrt{1-m}$. 

{\bf Solution XXXII}

Eqs. (\ref{A78}) and (\ref{A79}) also admit another periodic pulse solution 
\be\label{A94}
\phi = \frac{A\sqrt{m(1-m)}\sn(\beta y,m)}{\dn(\beta y,m)}\,, 
\ee
provided
\be\label{A95}
c_1 A^2 = -2\beta^2\,,~~c_3 = (2m-1)\beta^2\,.
\ee
Thus this solution is only valid if $c_1  < 0$ while $c_3 > (<)$ 0 depending
on if $m > (<) $1/2.

{\bf Solution XXXIII}

Remarkably, not only $1/\dn(y,m)$ and $\sn(y,m)/\dn(y,m)$ but even their superposition is an 
exact solution of Eqs. (\ref{A78}) and (\ref{A79}). In particular, it is easily
checked that
\be\label{A96}
\phi  = A\dn(\beta y,m) +\frac{B\sqrt{m(1-m)}\sn(\beta y,m)}{\dn(\beta y,m)}\,,
\ee
is also an exact solution of Eqs. (\ref{A78}) and (\ref{A79}) provided
\be\label{A97}
B^2 = A^2\,,~~2c_1 A^2 = -\beta^2\,,~~c_3 = -\frac{(1+m)\beta^2}{2}\,.
\ee
Thus this solution exists only if $c_1, c_3 < 0$.

{\bf Solution XXXIV}

Eqs. (\ref{A78}) and (\ref{A79}) also admit the periodic solution
\be\label{A98}
\phi = \frac{A\dn(\beta y,m)}{D+\sn(\beta y,m)}\,,~~ 0 < m < 1\,,~~ D > 1\,,
\ee
provided 
\be\label{A99}
2m c_1 A^2 = -(1-m)\beta^2\,,~~c_3 = \frac{(1+m)\beta^2}{2}\,, m D^2 = 1\,.
\ee

{\bf Solution XXXV}

Eqs. (\ref{A78}) and (\ref{A79}) also admit the periodic superposed 
solution
\be\label{A110}
\phi = \frac{A[\dn(\beta y,m)+B\sqrt{m}\cn(\beta y,m)]}
{D+\sn(\beta y,m)}\,,~~0 < m < 1\,,~~D > 1/\sqrt{m}\,,
\ee
provided 
\be\label{A101}
2b A^2 = -(D^2-1)\beta^2\,,~~2m b B^2 = -(mD^2 -1)\beta^2\,,~~c_3 = 
\frac{(1+m)\beta^2}{2}\,.
\ee

{\bf Solution XXXVI}

Eqs. (\ref{A78}) and (\ref{A79}) also admit a periodic kink solution
\be\label{A102}
\phi = \frac{A\sqrt{m}\sn(\beta y,m)}{1+B\cn^2(\beta y,m)}\,,~~B > 0\,,~~
0 < m < 1\,,
\ee
provided 
\be\label{A103}
B = \frac{\sqrt{m}}{1-\sqrt{m}}\,,~~c_1 A^2 = -8\sqrt{m} \beta^2\,,~~
c_3 = [6\sqrt{m}-(1+m)]\beta^2\,.
\ee

{\bf Solution XXXVII}

Eqs. (\ref{A78}) and (\ref{A79}) also admit a periodic pulse solution
\be\label{A104}
\phi = \frac{A\dn(\beta y,m)}{1+B\cn^2(\beta y,m)}\,,~~B > 0\,,~~
0 < m < 1\,,
\ee
provided 
\be\label{A105}
B = \frac{1-\sqrt{1-m}}{\sqrt{1-m}}\,,~~c_3 = [2-m+6\sqrt{1-m}]\beta^2\,,~~
c_1 A^2 = -\frac{8\beta^2}{\sqrt{1-m}}\,.
\ee

{\bf Solution XXXVIII}

Eqs. (\ref{A78}) and (\ref{A79}) also admit a periodic kink solution
\be\label{A106}
\phi = \frac{A\cn(\beta y,m)\sn(\beta y,m)}{1+B\cn^2(\beta y,m)}\,,
~~B > 0\,,~~ 0 < m < 1\,,
\ee
provided 
\be\label{A107}
B = \frac{1-\sqrt{1-m}}{\sqrt{1-m}}\,,~~ c_3 = [2-m -6\sqrt{1-m}]\beta^2\,,~~
c_1 A^2 = -\frac{8(1-\sqrt{1-m})^2 \beta^2}{\sqrt{1-m}}\,.  
\ee

{\bf Solution XXXIX}

Eqs. (\ref{A78}) and (\ref{A79}) also admit the periodic kink solution
\be\label{A108}
\phi = B + \frac{A\sn(\beta y,m)}{D+\sn(\beta y,m)}\,,~~ 0 < m < 1\,,~~ D > 1\,,
\ee
provided 
\be\label{A109}
A = -2 B\,,~~c_1 A^2 = -2(1-\sqrt{m})^2\beta^2\,,~~c_3 = 
\frac{[(1+m)+6\sqrt{m}]\beta^2}{2}\,, ~~\sqrt{m} D^2 = 1\,.
\ee

\end{document}